# 零边界条件下一维非线性细胞自动机可逆性的判定算法

马骏驰[1]　陈伟霖[1]　王晨[1]　林德福[1]　王超[1]

1 南开大学软件学院　天津　300350

**摘　要**　可逆性的性质对于经典的计算机科学理论模型——细胞自动机(CA)具有重要的意义。尽管 CA 在零边界条件下的线性规则的可逆性问题已经得到了大量的研究，但非线性规则目前还很少被探索。文中研究了在有限域 $\mathbb{Z}_p$ 上一般一维 CA 的可逆性问题，找到了一种优化 Amoroso 无限 CA 满射性判定算法的方法。基于此，文中还提出了在零边界条件下判定一维 CA 可逆性的算法，其中包括一种在零边界条件下判定一维 CA 严格可逆性的算法，以及一种基于桶链的在零边界条件下计算一维 CA 的可逆性函数的算法。这些判定算法不仅适用于线性规则，也适用于非线性规则。除此以外还证实了可逆性函数总是有一个周期的，且其周期性与对应桶链的周期性有关。文中给出了一些可逆 CA 的实验结果，并通过实验结果对理论部分进行了补充验证，进一步支持了文章的研究结论。

**关键词**：　细胞自动机；非线性规则；可逆性；零边界；一维

**中图法分类号**　（细化到 3 位数字）

# Decision algorithms for reversibility of one-dimensional non-linear cellular automata under null boundary conditions

MA Junchi[1]，CHEN Weilin[1]，WANG Chen[1]，LIN Defu[1] and WANG Chao[1]

1 College of Software，Nankai University，Tianjin

**Abstract**　The property of reversibility is quite meaningful for the classic theoretical computer science model, cellular automata. For the reversibility problem for a CA under null boundary conditions, while linear rules have been studied a lot, the non-linear rules remain unexplored at present. The paper investigates the reversibility problem of general one-dimensional CA on a finite field $\mathbb{Z}_p$, and proposes an approach to optimize the Amoroso's infinite CA surjectivity detection algorithm. This paper proposes algorithms for deciding the reversibility of one-dimensional CA under null boundary conditions. We propose a method to decide the strict reversibility of one-dimensional CA under null boundary conditions. We also provide a bucket chain based algorithm for calculating the reversibility function of one-dimensional CA under null boundary conditions. These decision algorithms work for not only linear rules but also non-linear rules. In addition, it has been confirmed that the reversibility function always has a period, and its periodicity is related to the periodicity of the corresponding bucket chain. Some of our experiment results of reversible CA are presented in the paper, complementing and validating the theoretical aspects, and thereby further supporting the research conclusions of this paper.

**Keywords**　cellular automata，non-linear rules，reversibility，null boundary，one-dimensional

## 1　引言

细胞自动机(CA)是由 Neumann[1]等人首先

到稿日期：　　　返修日期：
基金项目：天津市自然基金（21JCYBJC00210）
This work was supported by Tianjin Natural Science Fund (No. 21JCYBJC00210).
通信作者：王超（wangchao@nankai.edu.cn）

提出的一种理论计算机科学模型。在20世纪70年代，Conway设计了著名的"生命游戏"，其本质上是一个二维CA[2]。CA作为一种数学模型被学者们进行研究，并在各个领域得到了应用。时至今日，CA在复杂系统建模中的应用有车辆网络建模[3]、皮肤病模式建模[4]、建立肿瘤生长模型[5]、模拟行人疏散过程[6]等。在物理学中，CA还应用于磁场、电场等场的模拟[7]；在化学和材料学中，CA用于一系列材料效应的模拟，主要应用于再结晶、腐蚀现象[8]；在计算机科学中，CA可以被看作是并行计算机而用于并行计算的理论研究，同时CA也应用于图像处理[9]、图像加密[10]。可逆性是CA最重要的性质之一，因为一些应用场景要求系统具有可逆性，例如使用CA作为加密或编码工具的场景[11]。

最早被研究的是在空间中具有无穷多细胞的CA，我们称之为无限CA。满射问题被第一批研究CA的科学家称为伊甸园问题[12]。对于一维无限CA，Amoroso提出了一种确定满射性和单射性的方法[13]，Bruckner证明了如果无限CA是满射的，则它的局部变换函数必须是平衡的[14]。Sutner针对一维无限CA的满射性和单射性提出了一种二次时间复杂度的判定算法[15]。对于一维无限CA，单射性和可逆性已经被证明是等价的了。

对于具有有限个细胞的一维CA，我们称之为有限CA。由于不同的边界条件通常具有不同的性质，因此必须指定边界条件。迄今为止，研究最多的边界条件是零边界条件，它将所有边界外细胞的状态固定为零。Del Rey等人研究了初等线性CA的可逆性[16,17,18]。Sarkar讨论了零边界条件下混合初等CA的可逆性[19]。Cinkir研究了有限域上CA半径为2时的可逆性问题[20]。Yang讨论了一种在零边界条件下通过确定性有限自动机来判断一般的线性CA可逆性的方法[21]。Du在Yang的基础上进行了算法的优化[22]，提高了效率。到目前为止，关于零边界条件的结果大多局限于线性CA，而很少有理论或算法适用于非线性CA。

本文旨在研究零边界条件下一维CA的可逆性，并给出线性和非线性一维CA的判定算法。本文的其余部分组织如下：在第二节中我们做了一些初步的介绍，包括CA的基本理论和无限CA满射性的Amoroso判定过程；在第三节中，我们提出了零边界条件下一维CA可逆性的两种判定算法，一种是严格可逆性的判定算法，另一种是计算可逆性函数的算法，并给出了对应的实验结果。

## 2　基础理论

### 2.1　一维CA

一维CA是由一些数据单元组成的有限状态机，这些数据单元也被称为细胞。所有的细胞均匀的分布在一条直线上，且所有细胞的状态均在离散时间间隔内由局部规则同步变化。具体地说，可以用$(A, S, N, f)$这个四元组来定义一个一维CA，其中

- $A$代表一维CA中的动态细胞的集合，也称为细胞空间。通常，这些细胞是有序的，并通过数轴上的整数来匹配描述其坐标。
  - 如果$A$是无限集，则CA为无限CA，此处为$A = \mathbb{Z}$。
  - 如果$A$是一个包含$n$个细胞$(n \in \mathbb{Z}_+)$的有限集，则CA是一个有限CA。在这种情况下，$A = \{0, 1, 2, \ldots, n-1\} = \mathbb{Z}_n$。
- $S$表示状态集，包含细胞所有可能的状态。通常CA是在Galois素数域上定义的，即$S = \{0, 1, \ldots, p-1\} = \mathbb{Z}_p$，其中$p = |S| \geq 2$是素数。
- $N = (-r_L, \ldots, -1, 0, 1, \ldots, r_R)$表示邻域向量，

其中$r_L, r_R \geq 0$定义左半径和右半径，$k = r_L + r_R + 1$定义邻域大小。邻域定义了细胞之间的邻接关系。也就是说，下标为$i$的细胞其邻居细胞的下标为$i - r_L, ..., i, ..., i + r_R$。

- $f: S^k \mapsto S$表示局部规则，它将一个细胞的所有邻居当前时刻的状态映射到该细胞在下一个时刻的状态。
  - 如果$f$是线性组合的形式
  $$c_i^{t+1} = f(c_{i-r_L}^t, ..., c_i^t, ..., c_{i+r_R}^t) = \sum_{j=-r_L}^{r_R} \lambda_j c_{i+j}^t \bmod p \quad (1)$$
  其中$\lambda_j \in S$，$c_i^t \in S$表示下标为$i$的细胞在$t$时刻的状态，则该 CA 为一个线性 CA。
  - 否则该 CA 为一个非线性 CA。

注意，对于有限 CA，上述对于$f$的定义有一个缺陷，因为它缺少了对于边界以外的细胞状态的说明，即当$i < 0$或$i > n - 1$时$c_i^t$的取值。在本文中对于有限 CA，我们只考虑零边界条件。在零边界条件下，所有边界外细胞的状态恒定为0，这意味着

$$c_i^t \equiv 0, \quad \text{如果}i < 0 \text{ 或 } i > n - 1$$

对于其中各个$t$，如图1所示。

```
··· 0 | c₀ᵗ    c₁ᵗ    ···  cₙ₋₂ᵗ    cₙ₋₁ᵗ | 0 ···
                          ↓
··· 0 | c₀ᵗ⁺¹  c₁ᵗ⁺¹  ···  cₙ₋₂ᵗ⁺¹  cₙ₋₁ᵗ⁺¹ | 0 ···
```

图 1　零边界条件

Fig. 1　Null boundary conditions

所有细胞在$t$时刻的状态元组，表示为$C^t$，称为配置。在局部规则$f$的驱动下，全局函数$\Phi$表示从一个配置到下一时刻配置的映射，这意味着

$$\Phi(C^t) = C^{t+1}。$$

$C^t$称为$C^{t+1}$的原像。对于无限 CA，有全局映射函数$\Phi: S^\infty \mapsto S^\infty$；对于有限 CA，其中$C^t = (c_0^t, c_1^t, ..., c_{n-1}^t) \in S^n$，有全局映射函数$\Phi: S^n \mapsto S^n$。

如果 CA 的全局函数$\Phi$是双射，则该 CA 是可逆的。对于有限 CA 的$\Phi$，原像集和像集都是$S^n$，并且具有相同的基数，这意味着$\Phi$的双射性等价于它的单射性或满射性。我们有如下命题：

**命题 1**　有限 CA 是可逆的当且仅当任意$C \in S^n$恰好只有一个原像。

### 2.2　序列

为了方便描述，我们给出一些有关序列的定义。

序列是由有限个$S$中的元素组成的元组。如果元组长度为$l$，称其为一个$l$-序列。例如，对于有限一维 CA$(\mathbb{Z}_n, S, N, f)$，$t$时刻的配置$C^t = (c_0^t, c_1^t, ..., c_{n-1}^t)$是一个$n$-序列。局部规则$f$的原像集$S^k$中的每个元组是一个$k$-序列，其中$k$为邻域长度。

一个$l$-序列的一个后继是另一个$l$-序列，其最左$l - 1$个元素与原序列的最右$l - 1$个元素相同。对称地，一个$l$-序列的一个前驱是另一个$l$-序列，其最右$l - 1$个元素与原序列的最左$l - 1$个元素相同。例如，设$\alpha = (0,1,1,0,1)$，$\beta = (1,1,0,1,1)$是两个5-序列，或简写为$\alpha = 01101$，$\beta = 11011$，则$\beta$为$\alpha$的一个后继，$\alpha$为$\beta$的一个前驱。显然，每个序列恰有$|S|$个不同后继和$|S|$个不同前驱。

一个$l$-序列的后缀是将其最左元素移除后所得的$(l - 1)$-序列。对称地，一个$l$-序列的前缀是将其最右元素移除后所得的$(l - 1)$-序列。例如，对于前文中的序列$\alpha = 01101$，1101 是其后缀，0110 是其前缀。

### 2.3　Wolfram 规则号

在该部分，为了更好的解释 Wolfram 规则号，我们考虑状态集为$S = \mathbb{Z}_2$。在这种情况下，局部规则$f: \mathbb{Z}_2^k \mapsto \mathbb{Z}_2$可以通过（$Wolfram$）规则号$w^{[23]}$进行索引。邻域内的每个位置均可以用一个

二进制位来表示。最左边的位置对应最高位，最右边的位置对应最低位。这会将原像集$\mathbb{Z}_2^k$一一对应地映射到$I = \{i \in \mathbb{Z} | 0 \leq i \leq 2^k - 1\}$。然后，整数$i \in I$对应$w$中的第$i$位，该位对应于局部映射规则$f$所要映射的状态$a_i$，进而我们有该规则对应的Wolfram规则号$w = (a_{2^k-1}...a_2 a_1 a_0)_2 = \left(\sum_{i=0}^{2^k-1} a_i \cdot 2^i\right)_{10}$。例如，设置"$k = 3$"，则Wolfram规则号为$(150)_{10} = (10010110)_2$的局部规则映射表为

表1 Wolfram规则号为150的局部规则映射表

Table 1 Local rule mapping table with Wolfram rule number 150

| $(c_{i-1}^t, c_i^t, c_{i+1}^t)$ | $c_i^{t+1}$ | $(c_{i-1}^t, c_i^t, c_{i+1}^t)$ | $c_i^{t+1}$ |
|---|---|---|---|
| (0,0,0) | 0 | (1,0,0) | 1 |
| (0,0,1) | 1 | (1,0,1) | 0 |
| (0,1,0) | 1 | (1,1,0) | 0 |
| (0,1,1) | 0 | (1,1,1) | 1 |

### 2.4 Amoroso的判定过程

我们的算法是基于Amoroso[13]提出的一维无限CA的满射性判定算法，其执行步骤本质上是节点的构造，因此在执行完毕后我们可以将节点按序连接起来得到对应的Amoroso图。我们将在此介绍其构造算法。

**算法1**：Amoroso图构造及判定算法
**输入**：状态集$S$，局部规则$f$
**输出**：该CA的满射性

1:  **If** $\exists a \in S$，对任意$k$-序列都有$f(\alpha) \neq a$ **then**
2:      **return** 该CA是非满射的
3:  **end if**
4:  初始化Amoroso图$G = \{V, E\}$为空
5:  /* $V$为节点集，$E$为边集，后同此 */
6:  $b \leftarrow S$中的任一元素
7:  $X_{init} \leftarrow \{k - 序列\alpha \mid f(\alpha) = b\}$
8:  $V \leftarrow V \cup \{X_{init}\}$
9:  **foreach** $(X \in V, a \in S)$ **do**
10:     $X_a \leftarrow \{k - 序列\alpha \mid f(\alpha) = a$且$\alpha$是$X$中某序列的后继$\}$
11:     **if** $X_a = \emptyset$ **then**
12:         **return** 该CA是非满射的
13:     **end if**
14:     $V \leftarrow V \cup \{X_a\}$
15:     $E \leftarrow E \cup \{(X, X_a, a)\}$
16:     /* $(v_1, v_2, w)$：由$v_1$到$v_2$边权为$w$的有向边，后同此*/
17: **end foreach**
18: **return** 该CA是满射的

对该算法正确性的简要叙述如下。如果存在$a \in S$使得找不到$\alpha \in S^k$满足$f(\alpha) = a$，则任何包含一个或多个$a$的配置将找不到原像。因此，该CA是非满射的。

否则，每个配置对应Amoroso图上的一条无限长的路径，细胞状态对应边的标记。通过连续将节点中的序列与相邻节点中的前驱和后继连接（带有重叠），可以计算出该配置所有可能的原像。因此，如果某个节点内为空集，所有对应路径通过该节点的配置都必然不存在原像，因而该CA是非满射的。另一方面，如果每个节点都包含至少一个序列，则任意配置都至少存在一个原像，因而该CA是满射的。

由于不同$k$-序列的数量是有限的，不同节点的数量也是有限的。确切地说，不同节点的数量上界为$2^{(|S|^k)}$。一个节点内序列数量的上界为$|S|^k$。计算单个$k$-序列的后继的时间复杂度为$\theta(|S|)$。综上，算法1的最坏时间复杂度为$O(|S|^{k+1} \cdot 2^{(|S|^k)})$，此数值是一个极其宽松的上界，因为上述因子取最大值的情况微乎其微，并且不可能同时取最大值，平均情况远小于此。我们经过多组实验验证了上述分析。算法1的平均时间复杂度难以给出准确数值。

**例1** 假设$(\mathbb{Z}, \mathbb{Z}_2, \{-1,0,1\}, f)$为一维无限CA，其中$f$以规则号$(46)_{10} = (00101110)_2$为索引，由算法1构造的图如图2所示。具有空集的节点意味着...010...形式的所有配置都找不到原像。因此，规则号为46的CA是非满射的。

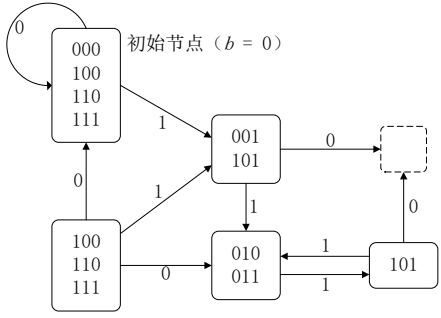

图 2 由规则号46构造的 Amoroso 图

Fig. 2 Amoroso graph constructed from rule number 46

## 3 零边界条件下一维 CA 的可逆性

### 3.1 Amoroso 图的构造及判定算法的优化

我们找到了一种优化 Amoroso 无限 CA 满射性判定算法的方法。因为本文的其余部分都会应用该优化，因此我们在此对其进行详细介绍。

简单地说，如果将图中所有节点内部的$k$-序列均替换为其后缀，修改后的 Amoroso 图依旧能判定对应 CA 的满射性。具体地，优化后的 Amoroso 图构造过程及判定算法如下。

**算法 2**：优化后的 Amoroso 图构造及判定算法
**输入**：状态集 $S$，局部规则 $f$
**输出**：该 CA 的满射性
1: If $\exists a \in S$，对任意$k$-序列都有$f(\alpha) \neq a$ then
2:    return 该 CA 是非满射的
3: end if
4: 初始化 Amoroso 图 $G = \{V, E\}$ 为空
5: $b \leftarrow S$ 中的任一元素
6: $X_{init} \leftarrow \{(k-1)-\text{序列}\gamma \mid \exists a \in S, f(a\gamma) = b\}$
7: $V \leftarrow V \cup \{X_{init}\}$
8: foreach $(X \in V, a \in S)$ do
9:    $X_a \leftarrow \{(k-1)-\text{序列}\gamma \mid \exists e \in S, f(e\gamma) = a \text{且} X \text{包含} e\gamma \text{的前缀}\}$
10:    if $X_a = \emptyset$ then
11:      return 该 CA 是非满射的
12:    end if
13:    $V \leftarrow V \cup \{X_a\}$
14:    $E \leftarrow E \cup \{(X, X_a, a)\}$
15: end foreach
16: return 该 CA 是满射的

对于此优化正确性的简要叙述如下。通过这样的替换，先前的$k$-序列被替换成其后缀，因而不会消失，这意味着非空集不会变为空集。另一方面，在替换前，对于图中每个节点，其后继节点的内容仅依赖于$k$-序列最右边的$k-1$个元素，因此替换后的$(k-1)$-序列不会改变构造后继节点过程中使用的任何信息。因此，替换前后的图是同构的（如果允许重复结点出现）。替换只会提高重复节点出现的可能性，例如：当$k = 3$时，001和101都被01取代，因此过程结束得更早（图变得更小，更简化）。

这种优化并不能节省构造单个节点的后继节点的复杂性。但是，原始版本的图的节点数量的上界为$2^{|S|^k}$，而优化版本将图的节点的上界优化为$2^{|S|^{k-1}}$。与算法1复杂度计算方式相同，算法2的最坏时间复杂度为$O(|S|^k \cdot 2^{(|S|^{k-1})})$，是算法1最坏时间复杂度的$\frac{1}{|S| \cdot 2^{|S|}}$。上述最坏复杂度是一个极其宽松的上界，平均情况远小于此。平均复杂度难以给出具体数值。

实验结果表明，对于本身在算法1中节点数较少的案例，算法2优化效果较低；而对于在算法1中节点数较多的案例，算法2优化效果十分显著。例如，对于直径为 3，状态集为$\mathbb{Z}_2$的情况，表2展示了在算法1中节点数较多的案例使用算法2的优化效果。

表 2 部分案例下算法 2 优化效果（$k = 3, S = \mathbb{Z}_2$）

Table 2 Efficiency of algorithm 2 for some cases（$k = 3, S = \mathbb{Z}_2$）

| 规则号（十进制） | 算法 1 节点数 | 算法 2 节点数 |
| --- | --- | --- |
| 22, 37, 104 | 23 | 16 |
| 41 | 18 | 13 |
| 85 | 11 | 5 |
| 86, 106 | 21 | 10 |
| 101 | 17 | 8 |
| 149, 169 | 23 | 11 |

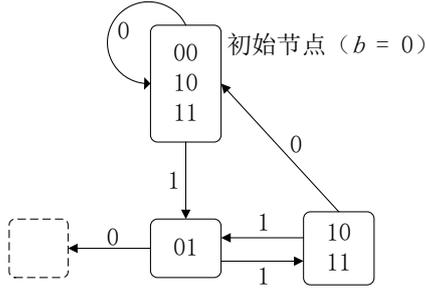

图 3　由规则号46构造的优化后的 Amoroso 图
Fig. 3　Optimized Amoroso graph constructed from rule number 46

**例 2**　同样考虑例1中的情况。在算法2中构造的图如图3所示。结果和以前一样，但这张图的尺寸很明显更小。

### 3.2　严格可逆性

对于一维有限 CA，其可逆性一般依赖于细胞数$n$。在一些情况下，我们会希望 CA 的可逆性并不会受到其细胞空间大小的影响，而一直保持可逆。下文中，细胞数$n$是一个变量。

**定义 1**　如果一维有限 CA $(\mathbb{Z}_n, S, N, f)$，对所有正整数$n$都可逆，则称为严格可逆。

**定义 2**　如果一个序列的最左侧$m$个元素均为0，则此序列是一个左$m$零序列。对称地，如果一个序列的最右侧$m$个元素均为 0，则此序列是一个右$m$零序列。

我们提出了一种在零边界条件下确定一维有限 CA 严格可逆性的算法，这也是一个构造图的过程。严格可逆性可通过算法3确定。

---
**算法 3**：严格可逆性的判定算法

**输入**：状态集$S$，邻域向量$(-r_L, ..., 0, ..., r_R)$，局部规则$f$
**输出**：该 CA 的严格可逆性
1:　初始化图$G = \{V, E\}$为空
2:　$X_{init} \leftarrow \{(k-1)-序列\gamma \mid \gamma为左r_L零序列\}$
3:　$V \leftarrow V \cup \{X_{init}\}$
4:　**foreach** $(X \in V, a \in S)$ **do**
5:　　$X_a \leftarrow \{(k-1)-序列\gamma \mid \exists e \in S, f(e\gamma) = a$且$X$包含$e\gamma$的前缀$\}$
6:　　**if** $X_a$不恰好包含一个右$r_R$零序列 **then**
7:　　　**return** 该 CA 不是严格可逆的
8:　　**end if**
9:　　$V \leftarrow V \cup \{X_a\}$
10:　　$E \leftarrow E \cup \{(X, X_a, a)\}$
11:　**end foreach**
12:　**return** 该 CA 是严格可逆的

---

图的构造与算法2非常相似。节点的后继节点以完全相同的方式构造。整个构造过程中只有两个不同之处：

● 初始节点中包含的$(k-1)$-序列。

在算法2中，初始节点包含所有在$f$下映射到同一$b \in S$的所有$k$-序列的后缀，因为对于无限 CA，配置是无限延伸的，初始节点必须对应到细胞空间内的某个细胞。

在算法3中，初始节点本身不对应细胞空间内的任何位置。对于零边界下的有限 CA，细胞空间内最左细胞的左侧均为恒定的 0 状态。我们希望初始节点的后继节点对应最左细胞，自然地，初始节点必须对应左边界外的第一个位置。因此，初始节点必须仅包含与零边界条件不冲突的序列，也就是左$r_L$零序列。换句话说，初始节点内的序列保证了在图上找到的所有原像都是符合左零边界的。

● 节点内的右$r_R$零序列。

在算法2中，CA 的非满射性由包含空集的节点判定，因为包含空集的节点意味着某些配置不存在原像。没有必要考虑任何特殊序列。

在算法3中，对于零边界下的有限 CA，只有适配零边界条件的原像才是有效的。如前文所提，原像与左侧零边界的适配由初始节点保证。对称地，原像与右侧零边界的适配由节点内的右$r_R$零序列保证。如果一个配置的原像以右$r_R$零序列作结尾，那么此原像适配右侧零边界。因此，若该 CA 是严格可逆的，则除初始节点外的每个节点必须包含恰好一个右$r_R$零序列。若一个节点不包含右$r_R$零序列，则某些配置不存在有效原像。若一个节点包含至少两个右$r_R$零序列，则某些配置存在多个有效原像。由命题1，上述任一情况都

意味着该 CA 的不可逆性。

图4给出了一个例子以更好地说明算法的原理。在此例中，$r_L = 2$，$r_R = 2$，$f$ 为规则号4161270000。路径对应配置 $(1,0,0,1,0)$，在节点中找到原像为 $(1,0,1,1,0)$。如图所示，原像与零边界的适配由初始节点中的序列 0010 以及尾节点中的序列 1000 保证。

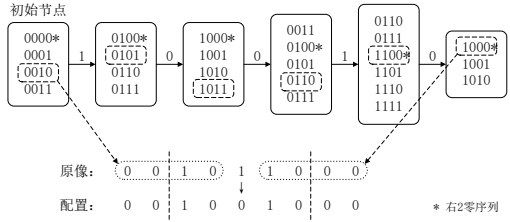

图 4　对于规则号 4161270000，算法3中图上的一条路径

Fig. 4　A path on the graph by Algorithm 3 for rule number 4161270000

此外，一旦出现第一个不包含恰好一个右 $r_R$ 零序列的节点（或包含空集的节点），严格可逆性就会被否定。因此，通常情况下并不需要构造整个图，在构造的过程中可以提前返回结果。然而，这一优化并不会降低算法的最坏时间复杂度。

算法 3 所使用图在构造时，除初始节点的设置外，其余与算法 2 均相同，故算法 3 的最坏时间复杂度为 $O(|S|^{r_L+r_R+1} \cdot 2^{(|S|^{r_L+r_R})})$，且同样地，这是一个很宽松的数值，平均情况远小于此。我们经过多组实验验证了上述分析。平均复杂度难以给出准确数值。

**例 3**　对于一维有限 CA$(\mathbb{Z}_n, \mathbb{Z}_2, \{-1,0,1\}, f)$，图5显示了算法3中构造的两张图，其中 $f$ 分别是通过规则号 $(153)_{10} = (10011001)_2$ 和规则号 $(150)_{10} = (10010110)_2$ 进行索引的。对于规则号153，所有节点均只包含一个右1零序列，因此 CA 是严格可逆的。对于规则号150，有一个节点不包含标记序列，而另一个节点有两个右1零序列，因此该 CA 不是严格可逆的。

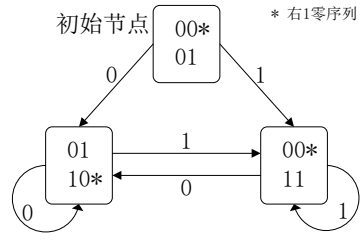

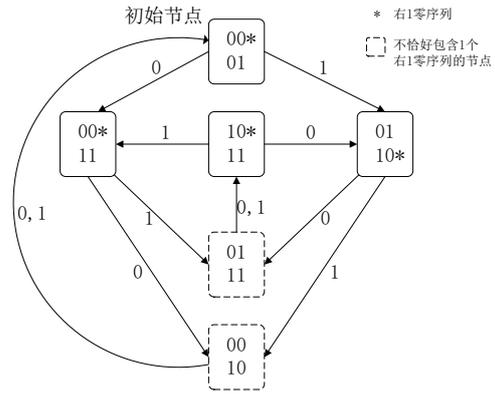

图 5　(a)规则编号153和(b)规则编号150的严格可逆性判定算法中的图

Fig. 5　The graph in the strictly reversible decision algorithm for (a) rule number 153 and (b) rule number 150

对于在 $\mathbb{Z}_2$ 上的一维有限 CA，我们利用算法得到了 $(r_L, r_R) = (1,1)$ 和 $(r_L, r_R) = (1,2)$ 时所有满足严格可逆性的规则，具体实验结果展示于表2和表3中。$(r_L, r_R) = (2,1)$ 的结果可以直接通过对称性从 $(r_L, r_R) = (1,2)$ 推导出来，因此我们省略了这一部分。实际上，我们还确认了当 $(r_L, r_R) = (2,2)$ 时，严格可逆的有186条，但是考虑到列举这些规则需要占用很大的篇幅，论文中不详细列出这些规则号。

表 3　满足严格可逆性的规则编号（$r_L = 1, r_R = 1$）

Table 3　Rule number that satisfies strict reversibility ($r_L = 1, r_R = 1$)

| 序号 | 十进制 | 二进制 |
| --- | --- | --- |
| 1 | 51 | 00110011 |
| 2 | 60 | 00111100 |

| 3 | 102 | 01100110 |
| 4 | 153 | 10011001 |
| 5 | 195 | 11000011 |
| 6 | 204 | 11001100 |

表 4 满足严格可逆性的规则编号（$r_L = 1, r_R = 2$）

Table 4 Rule number that satisfies strict reversibility ($r_L = 1, r_R = 2$)

| 序号 | 十进制 | 序号 | 十进制 |
|---|---|---|---|
| 1 | 3855 | 18 | 34695 |
| 2 | 3885 | 19 | 34725 |
| 3 | 3915 | 20 | 38550 |
| 4 | 4080 | 21 | 42375 |
| 5 | 7710 | 22 | 42405 |
| 6 | 7770 | 23 | 42465 |
| 7 | 11535 | 24 | 46260 |
| 8 | 11565 | 25 | 46320 |
| 9 | 15420 | 26 | 50115 |
| 10 | 19215 | 27 | 53970 |
| 11 | 19275 | 28 | 54000 |
| 12 | 23070 | 29 | 57765 |
| 13 | 23130 | 30 | 57825 |
| 14 | 23160 | 31 | 61455 |
| 15 | 26985 | 32 | 61620 |
| 16 | 30810 | 33 | 61650 |
| 17 | 30840 | 34 | 61680 |

### 3.3 可逆性函数

通过实验发现，在全体一维有限 CA 中，满足严格可逆性的是非常少的。通常情况下，一维有限 CA 对于部分细胞数$n$是可逆的，而对于另一些是不可逆的。可逆性函数描述了一维有限 CA 的这种性质。

**定义 3** 一维有限 CA $(\mathbb{Z}_n, S, N, f)$的可逆性函数用布尔函数$R$表示，其定义如下：

$$R(n) = \begin{cases} 1, & \text{如果该CA对于细胞数}n\text{是可逆的} \\ 0, & \text{如果该CA对于细胞数}n\text{是不可逆的} \end{cases}$$

特别的，$n \geq 1$，如果$R(n) \equiv 1$则一维有限 CA 是严格可逆的；如果$R(n) \equiv 0$则该一维有限 CA 是严格不可逆的。

**定义 4** 一个桶是由算法3中某些节点组成的集合。

桶中的节点是无序且不重复的。

我们提出了一种在零边界条件下计算一维有限 CA 的可逆性函数的算法，该算法是一个构造由一系列桶组成的"桶链"的过程。可逆性函数可通过算法4得到。

---

**算法 4**：可逆性函数的计算算法

**输入**：状态集$S$，邻域向量$(-r_L, ..., 0, ..., r_R)$，局部规则$f$

**输出**：该 CA 的可逆性函数$R$

1: $X_{init} \leftarrow \{(k-1) - 序列\gamma \mid \gamma \text{为左} r_L \text{零序列}\}$
2: $T_0 \leftarrow \{X_{init}\}, i \leftarrow 1, q \leftarrow -1$
3: **while** true **do**
4: $\quad T_i \leftarrow \emptyset$
5: $\quad$ **foreach** $(X \in T_{i-1}, a \in S)$ **do**
6: $\quad\quad X_a \leftarrow \{(k-1) - 序列\gamma \mid \exists e \in S, f(e\gamma) = a \text{且} X \text{包含} e\gamma \text{的前缀}\}$
7: $\quad\quad T_i \leftarrow T_i \cup \{X_a\}$
8: $\quad$ **end foreach**
9: $\quad R(i) \leftarrow 1$
10: $\quad$ **foreach** $X \in T_{i-1}$ **do**
11: $\quad\quad$ **if** $X = \emptyset$ **then**
12: $\quad\quad\quad R(i) \leftarrow 0$, break while
13: $\quad\quad$ **end if**
14: $\quad\quad$ **if** $X$不恰好包含一个右$r_R$零序列 **then**
15: $\quad\quad\quad R(i) \leftarrow 0$, break foreach
16: $\quad\quad$ **end if**
17: $\quad$ **end foreach**
18: $\quad$ **if** $\exists j \in [1, i-1], T_j = T_i$ **then**
19: $\quad\quad q \leftarrow i - j$, break while
20: $\quad\quad$ /*找到了一个循环*/
21: $\quad$ **end if**
22: $\quad i \leftarrow i + 1$
23: **end while**
24: $\forall j > i, R(j) \leftarrow \begin{cases} 0, & \text{if } q = -1 \\ R(j - q), & \text{if } q \neq -1 \end{cases}$
25: **return** $R$

---

算法4中使用桶的意图不难理解。算法3构造的图没有保存节点深度的信息，即每个节点与初始节点的距离。而一个配置的长度等于其对应路径的长度，即此条路径的末尾结点的深度。在算法4中，拥有相同深度的节点存放在同一个桶

中。桶$T_n$对应长度为$n$的路径。由于我们不关心具体每个配置的信息，原先图上的边和标签在算法4中被舍弃。用这种方式，一旦发现桶$T_n$中的某个节点不包含恰好一个右$r_R$零序列，就能立即判断该 CA 在细胞数为$n$时不可逆，即$R(n) = 0$。否则，如果桶$T_n$不包含这种节点，则该 CA 在细胞数为$n$时可逆，即$R(n) = 1$。

桶中每个节点是互不相同的，因而不同的桶的数量是有限的。因此，如果"桶链"足够长，必然会出现重复的桶。重复的两个桶之间"桶链"的结构在此之后无限循环，因此可逆性函数也必然在某个$n$之后具有周期性。

由可逆性函数的计算算法可知，对于一维有限 CA，其可逆性必然随细胞数的增长而周期性变化。此外，在此过程中，如果对某个$m$，我们在$T_m$中找到了包含空集的节点，则对所有$n \geq m$都必然有$R(n) = 0$，因为后续所有桶都必然有包含空集的节点。

关于算法 4 时间复杂度的分析如下：可以使用同算法 1 至算法 3 相似的方式给出一个极其宽松的最坏时间复杂度，这种方式给出的数值为$O\left(|S|^{r_L+r_R+1} \cdot 2^{2^{(|S|^{r_L+r_R})}}\right)$。然而，此数值毫无意义，因为我们的实验表明，没有任何一组用例能够达到甚至接近这个数值。例如在直径为 5 且状态集为$\mathbb{Z}_2$时，理论上算法 4 中不同桶的最大数量为 65536，然而我们的实验表明，在此情形下的所有用例中，桶的数量最多仅为 34。平均复杂度难以给出具体数值。

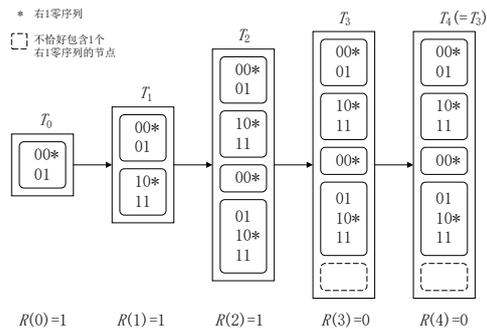

图 6　对于编号为19的规则的可逆性函数的计算过程

Fig.6 Computation of reversibility function for rule number 19

**例 4** 对于一维有限$CA(\mathbb{Z}_n, \mathbb{Z}_2, \{-1, 0, 1\}, f)$，其中$f$由规则号$(19)_{10} = (00010011)_2$表示，图6显示了其在零边界条件下的可逆性函数的计算过程。该过程在$T_4$处停止，因为$T_4$与$T_3$相同。因此，只有当细胞数等于1或2时，CA 才是可逆的。

对于状态集为$\mathbb{Z}_2$、邻域为$\{-1, 0, 1\}$的一维有限 CA，表3给出了零边界条件下所有对于某些$n \geq 2$可逆的规则号的可逆性函数的实验结果。结果也显示出对称性。

表 5　对于某些$n \geq 2$可逆的规则号的可逆性函数

Table 5 Reversibility function of rule numbers that are reversible for some $n \geq 2$.

($S = \mathbb{Z}_2$, $N = \{-1, 0, 1\}$)

| 规则号 | 可逆性函数 |
|---|---|
| $(5)_{10} = (00000101)_2$ | |
| $(37)_{10} = (00100101)_2$ | |
| $(122)_{10} = (01111010)_2$ | $R(n) = \begin{cases} 1, & n = 2 \\ 0, & \text{其他情况} \end{cases}$ |
| $(133)_{10} = (10000101)_2$ | |
| $(218)_{10} = (11011010)_2$ | |
| $(250)_{10} = (11111010)_2$ | |
| $(19)_{10} = (00010011)_2$ | |
| $(25)_{10} = (00011001)_2$ | |
| $(28)_{10} = (00011100)_2$ | |
| $(57)_{10} = (00111001)_2$ | |
| $(67)_{10} = (01000011)_2$ | |
| $(70)_{10} = (01000110)_2$ | |
| $(76)_{10} = (01001100)_2$ | |
| $(99)_{10} = (01100011)_2$ | $R(n) = \begin{cases} 1, & n = 1, 2 \\ 0, & \text{其他情况} \end{cases}$ |
| $(156)_{10} = (10011100)_2$ | |
| $(179)_{10} = (10110011)_2$ | |
| $(185)_{10} = (10111001)_2$ | |
| $(188)_{10} = (10111100)_2$ | |
| $(198)_{10} = (11000110)_2$ | |
| $(227)_{10} = (11100011)_2$ | |
| $(230)_{10} = (11100110)_2$ | |
| $(236)_{10} = (11101100)_2$ | |
| $(51)_{10} = (00110011)_2$ | |
| $(60)_{10} = (00111100)_2$ | |
| $(102)_{10} = (01100110)_2$ | $R(n) \equiv 1$ |
| $(153)_{10} = (10011001)_2$ | |
| $(195)_{10} = (11000011)_2$ | |
| $(204)_{10} = (11001100)_2$ | |
| $(90)_{10} = (01011010)_2$ | $R(n) = \begin{cases} 1, & n \equiv 0 \bmod 2 \\ 0, & 其他情况 \end{cases}$ |
| $(165)_{10} = (10100101)_2$ | |
| $(105)_{10} = (01101001)_2$ | $R(n) = \begin{cases} 1, & n \equiv 0,1 \bmod 3 \\ 0, & 其他情况 \end{cases}$ |
| $(150)_{10} = (10010110)_2$ | |
| $(108)_{10} = (01101100)_2$ | $R(n) = \begin{cases} 1, & n = 1,2,3 \\ 0, & 其他情况 \end{cases}$ |
| $(147)_{10} = (10010011)_2$ | |

## 结束语

本文研究了零边界条件下一维有限 CA 的可逆性。首先，受 Amoroso 图的构造及判定算法的启发，我们优化了该算法，降低了 Amoroso 图中节点数量的上界以提高运行效率。其次，我们提出了两种判定算法，它们对于线性和非线性 CA 均适用。第一种算法是构造一个图并根据图的特征来确定严格可逆性。第二种可逆性函数的计算算法是基于桶链得到的，其通过桶链结构的周期性确定 CA 对于任意细胞数的可逆性。使用这些算法，我们进行了大量的实验，并列出了一些规则号的结果。

对于可逆性问题，尽管目前人们对线性规则已经进行了大量的研究，但对于非线性规则的探索依旧很少。我们的算法最重要的贡献是提供了当 CA 的局部规则是非线性时确定可逆性的方法。现阶段我们的算法复杂度较高，难以在较大规模的 CA 中运行，因此我们下一步的工作重心将放在所提出算法的优化上。

## 参考文献

[1] Von Neumann J, Burks A W. Theory of self-reproducing automata[J]. IEEE Transactions on Neural Networks, 1966, 5(1): 3-14.

[2] Gardner M. Mathematical games[J]. Scientific american, 1970, 222(6): 132-140.

[3] Wang J , Lv W , Jiang Y ,et al. A cellular automata approach for modelling pedestrian-vehicle mixed traffic flow in urban city[J]. Applied Mathematical Modelling ,2023,115: 1-33.

[4] Kippenberger S, Bernd A, Thaçi D, et al. Modeling pattern formation in skin diseases by a cellular automaton[J]. The Journal of investigative dermatology, 2013, 133(2): 567.

[5] Valentim C A ,José A.Rabi, David S A .Cellular-automaton model for tumor growth dynamics: Virtualization of different scenarios[J].Computers in biology and medicine, 2022, 153:106481.

[6] Wang Y, Zhang L, Ma J, et al. Combining building and behavior models for evacuation planning[J]. IEEE computer graphics and applications, 2010, 31(3): 42-55.

[7] Ren F, Ge H, Fang H, et al. Simulation of the dendrite growth during directional solidification under steady magnetic field using three-dimensional cellular automaton method coupled with Eulerian multiphase[J]. International Journal of Heat and Mass Transfer, 2024, 218: 124809.

[8] Zhou F, Guo J, Zhao Y, et al. An improved cellular automaton model of dynamic recrystallization and the constitutive model coupled with dynamic recrystallization kinetics for microalloyed high strength steels[J]. Journal of Materials Research and Technology, 2023, Available online.

[9] Xu X, Fan C, Wang L. A deep analysis of the image and video processing techniques using nanoscale quantum-dots cellular automata[J]. Optik, 2022, 260: 169036.

[10] Darani A Y, Yengejeh Y K, Pakmanesh H, et al. Image encryption algorithm based on a new 3D chaotic system using cellular automata[J]. Chaos, Solitons & Fractals, 2024, 179: 114396.

[11] Cappellari L, Milani S, Cruz-Reyes C, et al. Resolution scalable image coding with reversible

cellular automata[J]. IEEE Transactions on Image Processing, 2010, 20(5): 1461-1468.

[12] Moore E F . Machine models of self reproduction, in proceedings of the fourteenth symposius on applied mathematics[J]. 1962, pp:17-33.

[13] Amoroso S, Patt Y N. Decision procedures for surjectivity and injectivity of parallel maps for tessellation structures[J]. Journal of Computer and System Sciences, 1972, 6(5): 448-464.

[14] Bruckner L K. On the Garden-of-Eden problem for one-dimensional cellular automata[J]. Acta Cybernetica, 1979, 4(3): 259-262.

[15] Sutner K. De Bruijn graphs and linear cellular automata[J]. Complex Systems, 1991, 5(1): 19-30.

[16] Del Rey A M. A note on the reversibility of elementary cellular automaton 150 with periodic boundary conditions[J]. Rom. J. Inf. Sci. Technol, 2013, 16(4): 365-372.

[17] Del Rey A M. A note on the reversibility of the elementary cellular automaton with rule number 90[J].Revista de la Unión Matemática Argentina, 2015, 56(1):107-125.

[18] Del Rey A M, Sánchez G R. On the reversibility of 150 Wolfram cellular automata[J]. International Journal of Modern Physics C, 2006, 17(07): 975-983.

[19] Sarkar P, Barua R. The set of reversible 90/150 cellular automata is regular[J]. Discrete applied mathematics, 1998, 84(1-3): 199-213.

[20] Cinkir Z, Akin H, Siap I. Reversibility of 1D cellular automata with periodic boundary over finite fields[J]. Journal of Statistical Physics, 2011, 143(4): 807-823.

[21] Yang B, Wang C, Xiang A. Reversibility of general 1D linear cellular automata over the binary field Z2 under null boundary conditions[J]. Information Sciences, 2015, 324: 23-31.

[22] Du X, Wang C, Wang T, et al. Efficient methods with polynomial complexity to determine the reversibility of general 1D linear cellular automata over Zp[J]. Information Sciences, 2022, 594: 163-176.

[23] Wolfram S, Gad-el-Hak M. A new kind of science[J]. Appl. Mech. Rev., 2003, 56(2): 18-19.

马骏驰，出生于 2000 年，硕士研究生，主要研究方向为细胞自动机。

陈伟霖，出生于 2000 年，硕士研究生，主要研究方向为细胞自动机。

王晨，出生于 2000 年，硕士研究生，主要研究方向为细胞自动机。

林德福，出生于 1996 年，硕士研究生，主要研究方向为细胞自动机。

王超，出生于 1976 年，博士，副研究员，天津市自然基金负责人，CCF 会员，主要研究方向为理论计算机科学，机器学习，人工智能。

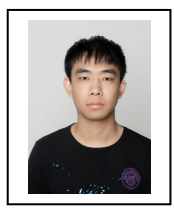

MA Junchi，born in 2000, postgraduate, is not a member of CCF. His main research interests include cellular automaton.

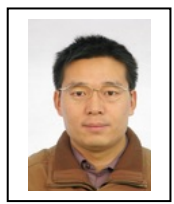

Wang Chao, born in 1976, associate researcher, a member of CCF. His main research interests is theoretical computer science, mainly cellular automata.